\documentclass[conference]{IEEEtran}
\usepackage[hyphens]{url}
\usepackage{hyperref}
\hypersetup{colorlinks,allcolors=blue}
\usepackage[backend=biber, style=ieee]{biblatex}

\usepackage{amsmath,amssymb,amsfonts}
\usepackage{algorithmic}
\usepackage{graphicx}
\usepackage{textcomp}
\usepackage{xcolor}
\usepackage{enumitem} 
\usepackage{booktabs}
\usepackage{orcidlink}  

\def\BibTeX{{\rm B\kern-.05em{\sc i\kern-.025em b}\kern-.08em
    T\kern-.1667em\lower.7ex\hbox{E}\kern-.125emX}}

\IEEEoverridecommandlockouts
\usepackage[font=footnotesize,justification=centering,labelsep=period,skip=2pt]{caption}

\begin{document}

\title{Advancing DevSecOps in SMEs: Challenges and Best Practices for Secure CI/CD Pipelines}

\author{\IEEEauthorblockN{Jayaprakashreddy Cheenepalli\,\orcidlink{0009-0006-0636-5432}, John D. Hastings\,\orcidlink{0000-0003-0871-3622}, Khandaker Mamun Ahmed\,\orcidlink{0000-0002-4713-188X}, Chad Fenner\,\orcidlink{0009-0000-1056-0456}}
\IEEEauthorblockA{\textit{The Beacom College of Computer and Cyber Sciences}\\
\textit{Dakota State University}\\
Madison, SD, USA \\
jp.cheenepalli@trojans.dsu.edu, \{john.hastings,khandakermamun.ahmed,chad.fenner\}@dsu.edu}
}

\maketitle

\begin{abstract}
This study evaluates the adoption of DevSecOps among small and medium-sized enterprises (SMEs), identifying key challenges, best practices, and future trends. Through a mixed methods approach backed by the Technology Acceptance Model (TAM) and Diffusion of Innovations (DOI) theory, we analyzed survey data from 405 SME professionals, revealing that while 68\% have implemented DevSecOps, adoption is hindered by technical complexity (41\%), resource constraints (35\%), and cultural resistance (38\%). Despite strong leadership prioritization of security (73\%), automation gaps persist, with only 12\% of organizations conducting security scans per commit. 

Our findings highlight a growing integration of security tools, particularly API security (63\%) and software composition analysis (62\%), although container security adoption remains low (34\%). Looking ahead, SMEs anticipate artificial intelligence and machine learning to significantly influence DevSecOps, underscoring the need for proactive adoption of AI-driven security enhancements. Based on our findings, this research proposes strategic best practices to enhance CI/CD pipeline security including automation, leadership-driven security culture, and cross-team collaboration. 

\end{abstract}


\section{Introduction}

Continuous integration and continuous delivery (CI/CD) \cite{fowler2006} pipelines have transformed software development by enabling frequent, automated releases. While this approach enhances speed and efficiency, it often overlooks security, leading to potential vulnerabilities that are not thoroughly tested \cite{b1}. Traditionally, security has been treated as a final stage concern, but this approach is inadequate given the increasing sophistication of cyber threats \cite{b2}. Industry research underscores the importance of early security integration in the development lifecycle, with studies showing that delayed security measures increase the likelihood of breaches compared to proactive security by design strategies \cite{b22,b23}. To address these challenges, DevSecOps integrates security throughout the CI/CD process, ensuring that security is embedded from the beginning rather than being an afterthought \cite{b3, b21}.

Unfortunately, for small to medium-sized enterprises (SMEs), adopting DevSecOps can present unique challenges due to limited resources, expertise, and infrastructure compared to larger organizations \cite{b4,b13}. The conventional approach of postponed security assessments often results in last minute security reviews, leading to heightened vulnerabilities and potential financial and reputational losses, particularly for SMEs \cite{b3}. Unlike large enterprises, SMEs often struggle with balancing agility and security while ensuring compliance with evolving regulatory requirements \cite{b12}. Without structured frameworks to guide security integration, SMEs often struggle with implementation, leading to gaps in security posture and compliance risks \cite{b5}. This research aims to bridge this gap by identifying tailored best practices that enhance security within SMEs' CI/CD pipelines \cite{b6}.

The research is guided by the following questions:

\begin{enumerate}[label={\textbf{RQ\arabic*:}},left=1.0em]
  \item What key challenges are faced by SMEs in adopting DevSecOps?

  \item What data-driven strategies and expert insights can help SMEs effectively implement DevSecOps?

    \item What are the best practices for DevSecOps adoption in SMEs?
\end{enumerate}

The remainder of the paper is organized as follows. Section \ref{meth} details the methodology for this study, followed by the results in Section \ref{results}. Section \ref{discuss} discusses the results, and Section \ref{recom} presents recommendations based on the findings. The paper ends with a discussion of future opportunities in Section \ref{limit} and the conclusion in Section \ref{conclude}.

\section{Methodology}\label{meth}

This research utilizes a mixed methods approach \cite{creswell2017} backed by a survey, integrating quantitative data analysis
with qualitative insights to thoroughly explore the challenges and strategies for implementing
DevSecOps in SMEs. The study is guided by two key theoretical frameworks:
\begin{itemize}

\item \textit{Technology Acceptance Model (TAM)}~\cite{davis1989perceived}: Assesses how perceived usefulness and ease of use influence SMEs' decisions to integrate security into DevSecOps. TAM guides the interpretation of survey responses regarding SMEs' attitudes towards adopting DevSecOps, shaping the questions to assess perceived usefulness and ease of implementation.

\item \textit{Diffusion of Innovations (DOI) Theory}~\cite{rogers2003diffusion}: Examines the stages of adoption, identifying barriers and drivers that influence DevSecOps implementation in SMEs. DOI provides a framework to interpret survey results about how SMEs adopt new practices, informing survey design and the subsequent qualitative analysis.

\end{itemize}

\begin{table*}[h!]
    \centering
    \caption{Key Elements of the Survey and their TAM/DOI Relevance}
    \label{tab:survey}
    \footnotesize
    \begin{tabular}{|p{5.15cm}|p{12.1cm}|}
    \hline
    \textbf{Survey Item} & \textbf{Theoretical Framework (TAM/DOI)} \\
    \hline
Organization Size & DOI: Examines how an organization's size influences DevSecOps adoption. \\
    \hline
    Role & TAM: Assesses perceived usefulness and ease of use of DevSecOps based on professional roles. \\
    \hline
    Duration in Current Role & TAM: Evaluates how experience level influences perceived ease of adoption. \\
    \hline
    Current Implementation Status & DOI: Identifies the organization's DevSecOps adoption stage within the innovation lifecycle. \\
    \hline
    Security Tools Integration (CI/CD) & TAM: Analyzes how SMEs perceive the usefulness and ease of integrating security tools into CI/CD. \\
    \hline
    Security Tools Integration in IDEs & TAM: Measures how seamlessly security tools are integrated into development environments and their perceived value. \\
    \hline
    Primary Challenges & TAM \& DOI: Identifies barriers such as complexity, resource constraints, and cultural resistance, influencing adoption difficulty. \\
    \hline
    Factors Supporting Adoption & DOI: Highlights drivers such as leadership support, compatibility \& org. readiness for DevSecOps. \\
    \hline
    Effectiveness in Security Improvement & TAM: Assesses perceived benefits of DevSecOps in improving security, monitoring \& vulnerability mgmt. \\
    \hline
    Priority of Security by Leadership & DOI: Evaluates leadership’s role in promoting security as a strategic priority and its impact on adoption. \\
    \hline
    Integration of Security Practices & TAM \& DOI: Measures the extent of security integration in development workflows and its visibility within teams. \\
    \hline
    Frequency of Security Practices in Workflows & TAM \& DOI: Examines the regularity of security processes, influencing their perceived ease of adoption and observability. \\
    \hline
    Metrics for Effectiveness & TAM: Identifies key performance indicators (KPIs) used to measure the success of DevSecOps adoption. \\
    \hline
    Frequency of Security Training & DOI: Assesses the role of training frequency in improving security culture and trialability. \\
    \hline
    Additional Resources Required & DOI: Identifies resource gaps (budget, tools, training) necessary to facilitate smoother adoption. \\
    \hline
    Desired Improvements & DOI: Highlights areas SMEs wish to enhance for optimizing DevSecOps effectiveness. \\
    \hline
    Impact of Early Security Integration & TAM: Evaluates how embedding security early in DevSecOps improves software quality \& risk mitigation. \\
    \hline
    Ease of Implementing DevSecOps & TAM: Measures perceived difficulty in adopting DevSecOps processes \& maintaining security automation. \\
    \hline
    Frequency of Security Scans & TAM \& DOI: Tracks scan frequency \& its influence on automation adoption, ease of use \& observability.\\
    \hline
    Difficulty of Integration into CI/CD Pipeline & TAM: Identifies perceived technical complexity in embedding security within development workflows. \\
    \hline
    Cross-team Collaboration & DOI: Evaluates teamwork, communication \& integration between security, development \& operations. \\
    \hline
    Organizational Engagement with DevSecOps & DOI: Determines where SMEs stand on the DevSecOps maturity scale, from laggards to early adopters. \\
    \hline
    Challenges in Adoption & DOI: Analyzes qualitative insights on barriers preventing smoother DevSecOps adoption. \\
    \hline
    Anticipated Technological Advancements & TAM \& DOI: Examines expectations for AI and automation in enhancing future DevSecOps efficiency. \\
    \hline
\end{tabular}
\end{table*}

By leveraging these frameworks, this research aims to provide structured insights into both the technical and organizational challenges that SMEs face in DevSecOps adoption. The goal is to propose practical, scalable best practices aligning with SMEs' operational needs, security constraints, and business objectives, ensuring security is seamlessly integrated into development workflows~\cite{b6,b27,b28}. 

This methodology effectively addresses both the practical and theoretical components of DevSecOps, ensuring robust and applicable
findings \cite{b23}. By integrating empirical data with expert insights, the study provides a
solid framework for identifying best practices and potential pitfalls, facilitating the development
of effective strategies for enhancing DevSecOps \cite{b28,b17}.

\subsection{Data Collection}

Data were collected using a survey instrument organized according to the items shown in Table \ref{tab:survey} with the relevance of TAM and DOI for each item shown. 

\subsection{Survey Distribution and Response Rate}
Participants were recruited through Centiment \cite{centiment}, a secure survey distribution service specializing in targeted respondent recruitment and anonymous data collection. Centiment fully managed the survey distribution process, providing respondents who met specific eligibility criteria: participants had to be professionals working in SMEs with experience in DevSecOps and actively involved in software development, operations, or security. Individuals without direct experience in DevSecOps were excluded. Participants accessed the survey via secure links provided by Centiment at their convenience. 

\subsection{Ethical Considerations}

Due to the sensitive nature of
security data, participant anonymity and strict data protection
measures were prioritized. The study received Institutional Review Board approval under DSUIRB-20241202-11, confirming compliance with ethical standards.

\section{Results}\label{results}

The survey method resulted in 405 valid responses. The following sections present a detailed analysis of the key findings from these responses, categorized into nine themes.

\subsection{DevSecOps Implementation Maturity}

Table \ref{tab:adopt_levels} presents the maturity levels of DevSecOps adoption among SMEs. The majority (68\%) of respondents have fully implemented DevSecOps practices, demonstrating a strong commitment to security integration. Additionally, 22\% of SMEs are in the process of adopting these practices, while 10\% have not yet started implementation. These results emphasize the increasing recognition of DevSecOps as a vital component of SME security strategies, reflecting a positive trend toward enhanced security practices in the industry.

\begin{table}[h]
    \centering
    \caption{DevSecOps Adoption Levels Among SMEs}
    \label{tab:adopt_levels}
    \begin{tabular}{l c}
        \toprule
        \textbf{Adoption Level} & \textbf{Percentage} \\
        \midrule
        Fully Implemented & 68\% \\
        In the Process of Adoption & 22\% \\
        Not Yet Started & 10\% \\
        \bottomrule
    \end{tabular}
\end{table}

\subsection{Roles, Organization Size, and Experience}

\begin{figure}[h]
    \centering
    \includegraphics[width=0.45\textwidth]{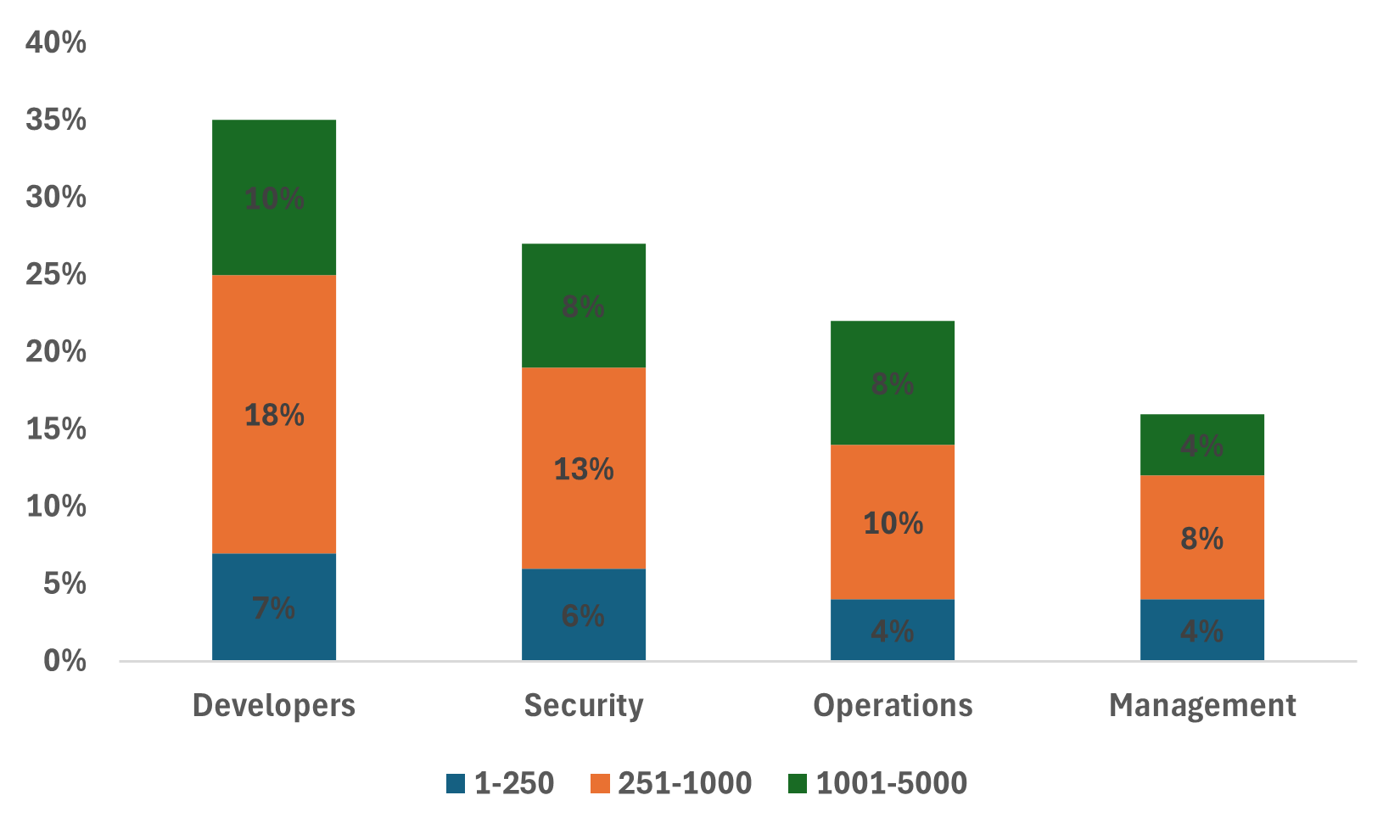} 
    \vspace{-4pt}
    \caption{Roles Distribution Across SME Sizes}
    \label{fig:roles}
\end{figure}

The study predominantly represents SMEs, ensuring relevance to this segment. As seen in Fig. \ref{fig:roles}, developers (35\%) are the largest group, followed by security (27\%), operations (22\%), and management (16\%), highlighting cross-functional involvement in DevSecOps adoption. 

In terms of experience, 49\% of respondents have over five years in the industry, with 81\% having more than three years, ensuring well-informed insights. These findings provide a strong foundation for analyzing DevSecOps maturity and adoption trends.

\subsection{Security Tool Adoption in Development Pipelines}

\begin{figure}[h]
    \centering
    \includegraphics[width=0.5\textwidth]{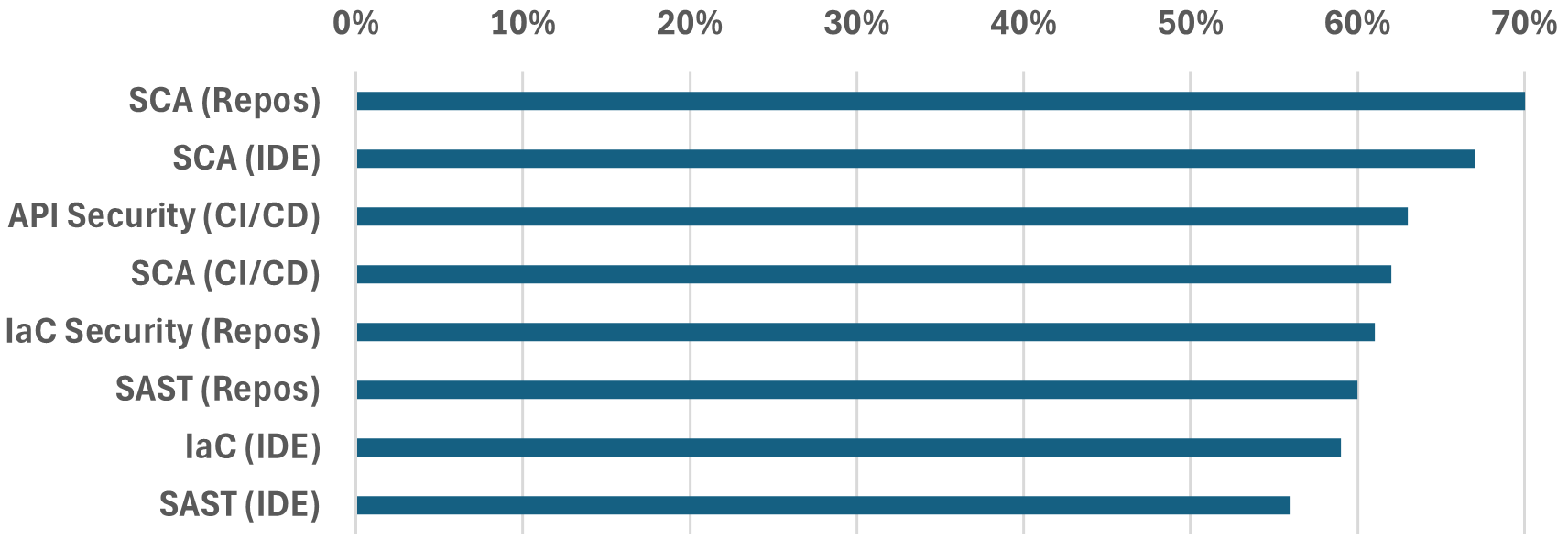} 
    \caption{Security Tool Adoption in Development Pipelines}
    \label{fig:tool}
\end{figure}

As seen in Fig. \ref{fig:tool}, the study found strong security integration in development pipelines, with high adoption of API Security (63\%), SCA (62\%), IaC Security (61\%), and SAST (60\%) in CI/CD and source code repositories. IDE integration is most common for SCA (67\%), followed by IaC (59\%) and SAST (56\%), reinforcing developer-centric security practices. However, low container security adoption (34\%) indicates a gap in securing containerized environments. These findings emphasize both progress and areas for improvement in SME security strategies.

\subsection{Primary Challenges in DevSecOps Adoption}

\begin{figure}[h]
    \centering
    \includegraphics[width=0.40\textwidth]{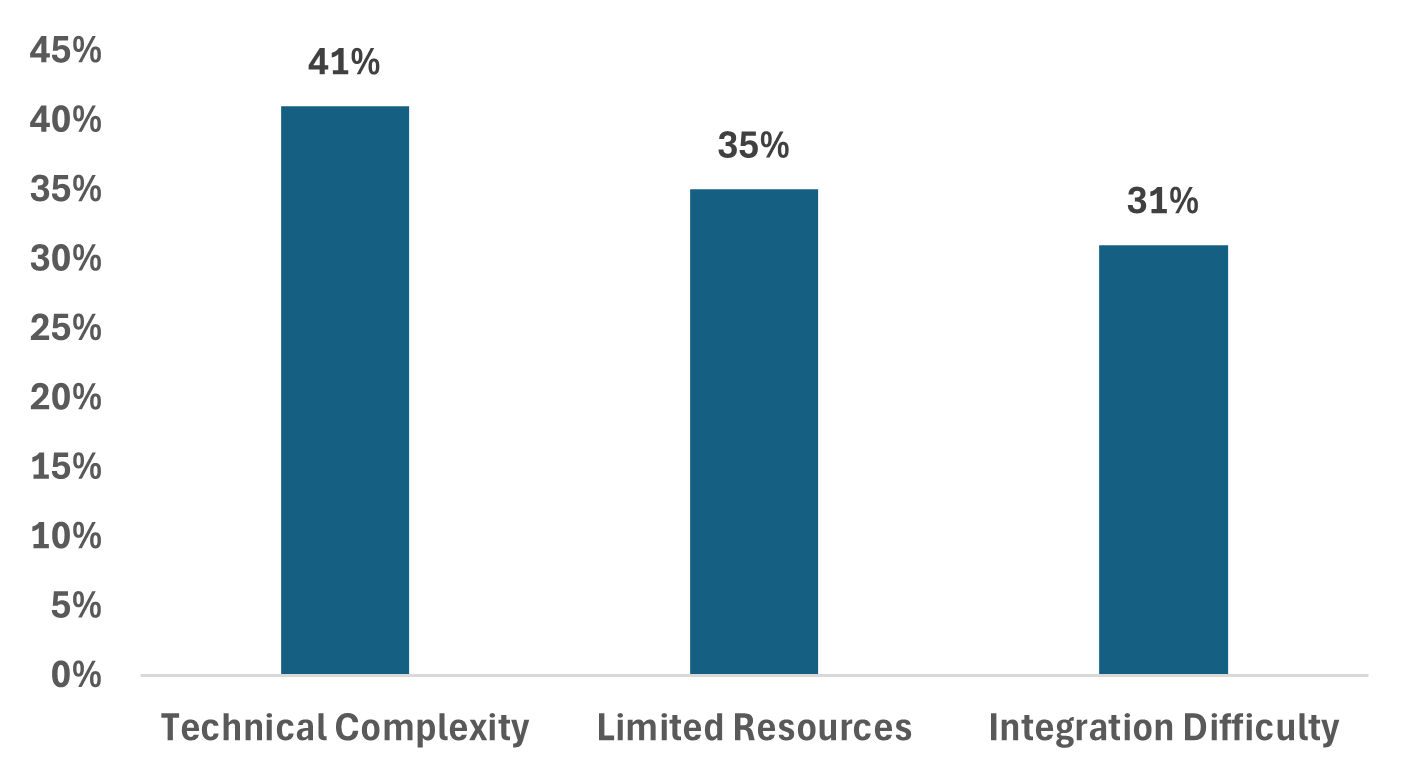} 
    \caption{Primary Challenges in DevSecOps Implementation}
    \label{fig:challenge}
\end{figure}

As shown in Fig. \ref{fig:challenge}, the study identified technical complexity (41\%) and limited resources (35\%) as the top barriers to DevSecOps adoption among SMEs. These challenges highlight the need for simplified tools, better automation, and increased resource allocation to facilitate seamless integration.

Additionally, 31\% of respondents reported moderate difficulty integrating security into CI/CD pipelines, indicating persistent hurdles in embedding security practices early in the development lifecycle. These findings emphasize the necessity of improved tooling, automation, and streamlined processes to enhance DevSecOps adoption.

\subsection{Effectiveness and Metrics of DevSecOps Practices}

As seen in Table \ref{tab:effect}, 67\% of SMEs perceive DevSecOps as highly effective, correlating with improved security monitoring, vulnerability management, and incident reduction. These findings reinforce the value of integrating security practices throughout the development lifecycle. 
Key success metrics include the frequency of security scans (53\%), vulnerabilities detected (46\%), remediation time (45\%), and reduced security incidents (44\%). These metrics emphasize the importance of proactive security monitoring and timely response measures. 

\begin{table}[h]
    \centering
    \caption{Effectiveness vs. Key Metrics in DevSecOps Practices}
    \label{tab:effect}
    \begin{tabular}{l c}
        \toprule
        \textbf{Metric} & \textbf{Percentage} \\
        \midrule
        Effectiveness & 67\% \\
        Frequency of Security Scans & 53\% \\
        Vulnerabilities Detected & 46\% \\
        Remediation Time & 45\% \\
        Reduced Incidents & 44\% \\
        \bottomrule
    \end{tabular}
\end{table}

\subsection{Leadership Priority and Collaboration Levels in DevSecOps}

As shown in Table \ref{tab:leader}, leadership commitment plays a crucial role in DevSecOps adoption, with 73\% of respondents rating security as a high or extremely high priority. This strong focus on security governance reinforces the importance of top-down prioritization in driving successful implementations.

\begin{table}[h]
    \centering
    \caption{Leadership Priority and Collaboration Levels in DevSecOps}
    \label{tab:leader}
    \begin{tabular}{l c}
        \toprule
        \textbf{Category} & \textbf{Percentage} \\
        \midrule
        High or Extremely High Leadership Priority on Security & 73\% \\
        Cross-Team Collaboration Encouraged & 69\% \\
        Technological Readiness as a Driver & 66\% \\
        Leadership Support as a Driver & 43\% \\
        Cultural Resistance as a Challenge & 38\% \\
        \bottomrule
    \end{tabular}
\end{table}

Additionally, cross-team collaboration is actively encouraged by 69\% of organizations, emphasizing the need for seamless cooperation between development, security, and operations teams. However, cultural resistance (38\%) remains a significant challenge, highlighting the necessity for ongoing cultural transformation initiatives to align teams with security goals. Furthermore, technological readiness (66\%) and leadership support (43\%) were identified as key drivers of DevSecOps adoption. These findings underscore the need for leadership-driven initiatives and investment in technology to foster a security-first mindset.

\subsection{Training, Education, and Resource Needs}

\begin{figure}[h]
    \centering
    \includegraphics[width=0.45\textwidth]{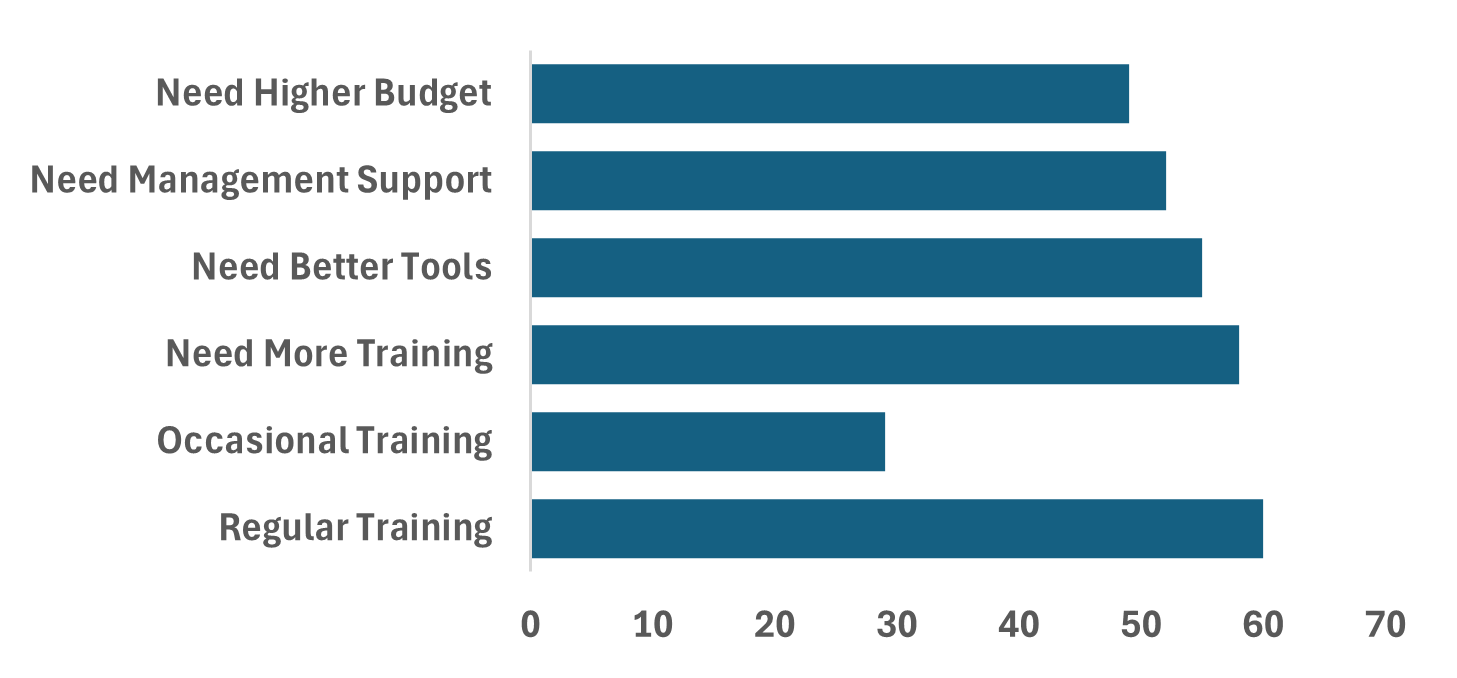} 
    \vspace{-2pt}
    \caption{Regular and Occasional Training, and Additional Resources Needed}
    \label{fig:train}
\end{figure}

As shown in Fig. \ref{fig:train}, 60\% of SMEs provide regular security training, while 29\% offer it only occasionally, revealing gaps in consistent training efforts. 
There is a strong demand for additional resources, with 58\% requesting enhanced training, 55\% seeking better security tools, 52\% needing stronger management support, and 49\% requiring increased budget allocations. These findings emphasize the need for improved training programs and resource investment to optimize DevSecOps implementation.

\subsection{Anticipated Future Technological Impacts}

\begin{figure}[h]
    \centering
    \includegraphics[width=0.45\textwidth]{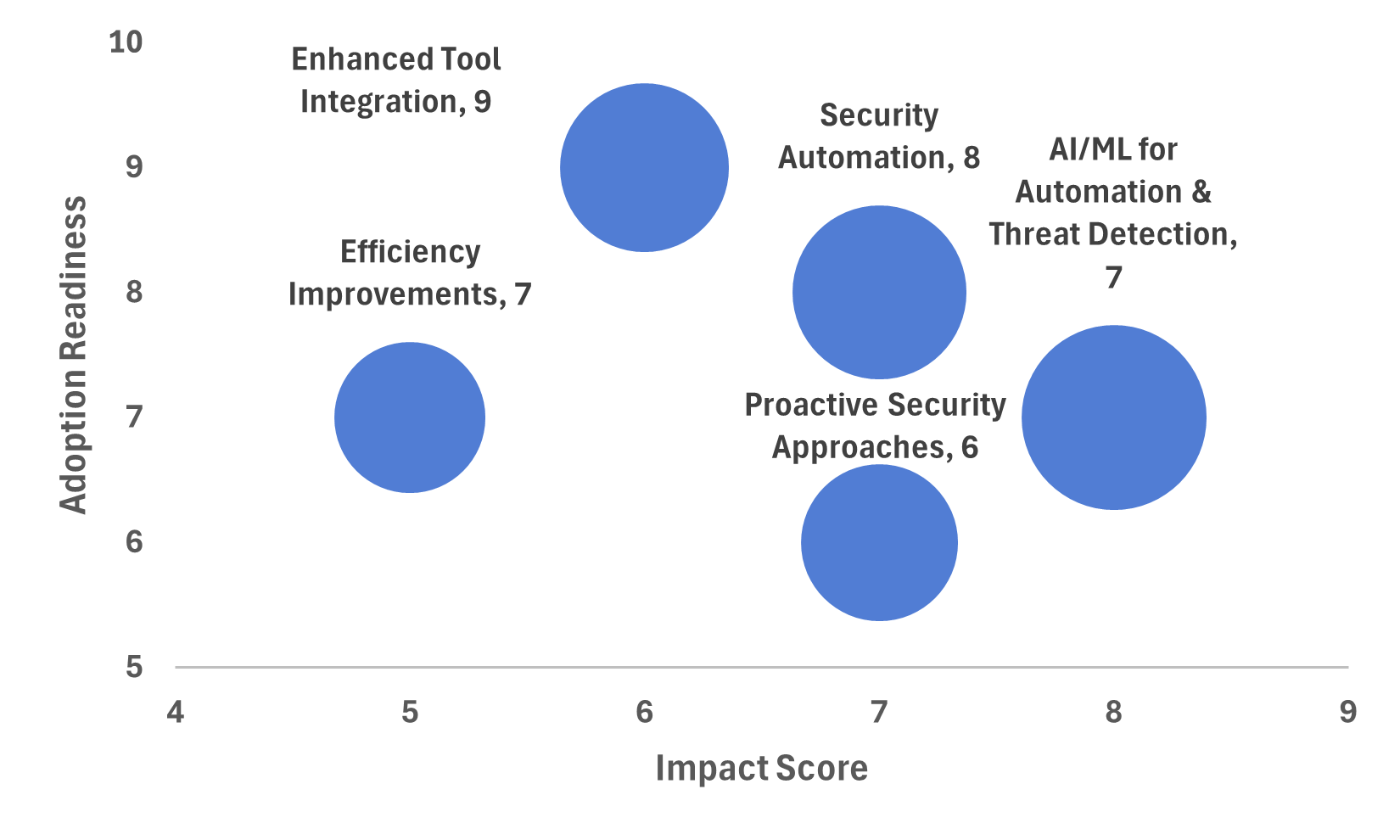} 
    \vspace{-6pt}
    \caption{Expected Technological Advancements Impact} 
    \label{fig:impact}
\end{figure}

As seen in Fig. \ref{fig:impact}, SMEs anticipate that AI and machine learning (ML) will significantly impact DevSecOps over the next 2-3 years, particularly in automation, threat detection, and enhanced tool integration. These advancements are expected to drive real-time vulnerability detection, proactive security approaches, and operational efficiency improvements. 
The findings suggest a shift towards AI-driven security strategies, emphasizing the need for continuous skill development and integration of intelligent security solutions to streamline DevSecOps practices.

\subsection{Operational Integration and Early Adoption of Security Practices}

\begin{figure}[h]
    \centering
    \includegraphics[width=0.45\textwidth]{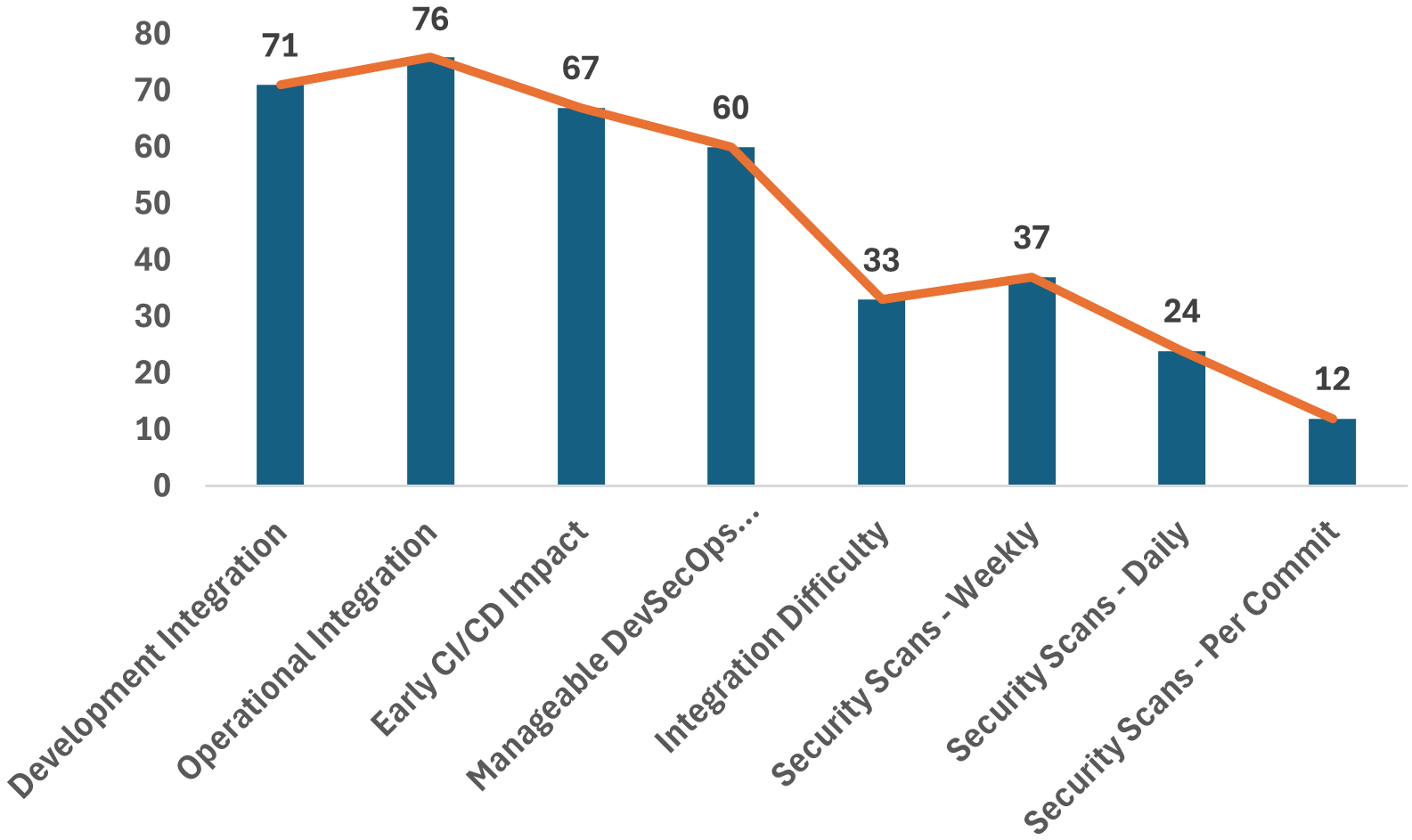} 
    \caption{Operational Integration and Early Adoption of Security Practices}
    \label{fig:integ}
\end{figure}

As shown in Fig. \ref{fig:integ}, the study found strong security integration into both development (71\%) and operational workflows (76\%), reinforcing DevSecOps as a key aspect of operations. Early security integration in CI/CD pipelines (67\%) is perceived to improve software quality, validating its importance in secure development. 
Although 60\% of SMEs find DevSecOps manageable to implement, 33\% report moderate integration difficulty, suggesting the need for improved automation and streamlined processes. In addition, security scans occur weekly (37\%) or daily (24\%), but only 12\% scan per commit, indicating room for enhanced automated security practices.

\section{Discussion}\label{discuss}

The findings of this study reveal the significant strides SMEs have made in adopting DevSecOps, with a clear majority already fully implementing or making substantial progress towards integration. This high adoption rate reflects the growing recognition among SMEs of DevSecOps' critical role in enhancing security, efficiency, and overall business resilience. The prevalence of SMEs reporting themselves as early adopters or leaders further indicates a mature understanding of the value that DevSecOps brings to modern software development environments.

Despite widespread adoption, the survey highlights notable challenges hindering broader implementation. Technical complexity remains the foremost obstacle, suggesting that current DevSecOps tools and processes might be too intricate or not sufficiently tailored to SME needs. Coupled with limited resources and expertise gaps, these challenges pose considerable barriers to full-scale adoption. The relatively lower adoption of container security practices indicates a specific area requiring attention, as containerized applications become increasingly prevalent in software development.

Leadership commitment emerges as a pivotal driver in successful DevSecOps practices. SMEs with high leadership engagement reported significantly better integration of security practices, cross-team collaboration, and higher overall effectiveness of their security posture. Conversely, organizations experiencing cultural resistance reported slower adoption rates, underscoring the necessity of strong leadership and a supportive organizational culture to ensure successful DevSecOps implementation.

Regular security training is prevalent among SMEs; however, the expressed need for enhanced training, better tools, stronger leadership support, and increased budgets indicates that current practices are often insufficient to sustain optimal security practices. Addressing these resource gaps through strategic investments in targeted training programs and automation tools could greatly enhance the effectiveness and sustainability of DevSecOps practices.

Anticipated advancements in AI and ML technologies represent substantial opportunities for SMEs to overcome existing challenges. Respondents expect these technologies to automate processes, improve threat detection, and streamline vulnerability management, potentially mitigating many current barriers identified in the survey. Nevertheless, the readiness to adopt these advancements varies among SMEs, highlighting a need for continued education and resource allocation to leverage these emerging technologies effectively.

The integration of security into both development and operational workflows appears robust, reinforcing that SMEs recognize security as integral rather than supplementary. Regular security scans, particularly weekly and daily, indicate a positive shift towards proactive vulnerability management. However, opportunities remain to further optimize the frequency and automation of security checks, especially integrating them into every code commit, a practice currently less common.

\section{Recommendations}\label{recom}
Based on the findings, this research proposes the following best practices tailored to SMEs adopting DevSecOps: 

\begin{enumerate}
\item \textit{Foster a security first culture}:
A security first culture requires strong executive advocacy and the allocation of dedicated security resources. Establish regular security awareness programs to ensure that security is embedded into daily operations, fostering a proactive approach. Additionally, defining clear accountability across teams helps maintain security as a shared responsibility rather than an isolated function.

\item \textit{Seamlessly integrate tools and automation}:
Integrate security tools into development environments, including IDEs, repositories, and CI/CD pipelines. Automating frequent security scanning and vulnerability remediation enhances efficiency and reduces manual errors. 

\item \textit{Enhance cross team collaboration}:
Effective DevSecOps adoption requires strong collaboration between development, security, and operations teams. Establishing cross-functional teams improves communication and fosters shared responsibility for security. Utilizing collaborative platforms and holding regular meetings ensure better alignment and visibility across all stakeholders.

\item \textit{Continuous training}:
Providing structured, security-focused training and certification programs is essential for maintaining an informed workforce. Encouraging skill sharing and mentorship allows employees to continuously develop their security expertise. Tailored onboarding programs for DevSecOps processes help new employees quickly acclimate to security best practices.

\item \textit{Executive leadership engagement}:
Active participation by leadership is critical for the success of DevSecOps initiatives. Conducting leadership workshops reinforces the importance of security within business operations. Aligning DevSecOps initiatives with overarching business goals ensures sustained commitment and resource allocation from executives.

\item \textit{Proactive regulatory compliance management}:
Embedding compliance checks within development workflows ensures that security and regulatory requirements are met from the outset. Automating regulatory monitoring and reporting reduces the burden on security teams and enables proactive compliance management.

\item \textit{Strategic budget and resource management}:
Adopting a phased approach to DevSecOps implementation allows organizations to achieve incremental success without overwhelming resources. Prioritizing cost effective and open source solutions helps SMEs maximize security investments while maintaining financial sustainability.

\item \textit{Leverage AI and ML}:
Integrate AI-driven solutions to enhance security operations by enabling real-time threat detection and automated incident response. Providing specialized training in AI enhanced security tools ensures that teams can effectively utilize these technologies to strengthen DevSecOps processes.

\item \textit{Enhance Infrastructure as Code (IaC)}:
Regular auditing of IaC templates and configurations helps maintain security and compliance within infrastructure automation. Automating security policy enforcement ensures that best practices are consistently applied across all cloud and on premise environments.

\item \textit{Robust metrics for continuous improvement}:
Define and track key performance indicators (KPIs), such as scan frequency and vulnerability remediation rates, to measure and optimize DevSecOps effectiveness. Regularly reviewing and refining DevSecOps processes based on collected metrics ensures continuous improvement and adaptation to evolving threats.

\item \textit{Promote cultural change}:
Promote a security conscious culture through sustained efforts, including internal campaigns and reward programs for security champions. Gradually embedding DevSecOps practices across teams helps reinforce security as a fundamental aspect of the development lifecycle and operational processes.
\end{enumerate}

\section{Future Work}\label{limit}
While this research provides comprehensive insights into DevSecOps adoption among SMEs, it also presents several opportunities for further exploration. First, the study primarily focuses on SMEs within a specific range of employee sizes (1-5000 employees), which may limit the generalizability of findings to larger enterprises. Future research should explore broader organizational sizes to determine whether the proposed best practices remain applicable across varying business scales.

Second, this study relies on self-reported survey data, which may introduce biases such as overestimation or underestimation of DevSecOps adoption levels and effectiveness. Future work should incorporate empirical validation techniques, such as case studies, expert interviews, or direct observational data, to corroborate the findings and strengthen the reliability of conclusions.

Third, while this study identifies key technological advancements such as AI and ML as future enablers of DevSecOps, it does not explore their practical implementation in depth. Future research should conduct experimental studies to assess the impact of AI-driven security automation and predictive analytics in real-world DevSecOps environments.

Additionally, the research does not examine industry-specific DevSecOps challenges. SMEs in highly regulated industries such as finance and healthcare may face distinct security and compliance hurdles compared to technology startups. Future work should conduct sector-specific analyses to tailor best practices accordingly.

Finally, as the cybersecurity landscape evolves, this study provides a snapshot of current DevSecOps trends and challenges. Future research should revisit and update these findings periodically to reflect emerging threats, evolving best practices, and the latest advancements in security automation. Expanding the study to include longitudinal research could provide deeper insights into the long-term effectiveness of DevSecOps adoption and organizational security postures.

\section{Conclusion}\label{conclude}

This research provides comprehensive insights into DevSecOps adoption among SMEs, highlighting both the maturity of existing implementations and persistent challenges such as technical complexities, budget constraints, and skill gaps. The study’s findings reinforce the importance of integrating security seamlessly into CI/CD pipelines and adopting a security first mindset to mitigate emerging cybersecurity risks.

The proposed best practices serve as a practical roadmap for SMEs, offering actionable strategies to strengthen security integration, enhance cross-team collaboration, and improve operational resilience. By aligning these practices with the TAM and DOI theory, this research ensures that the recommendations are not only technically viable but also culturally and organizationally adaptable.

Moreover, the increasing role of AI and ML in DevSecOps presents an opportunity for SMEs to enhance automation, optimize security operations, and proactively mitigate threats. As organizations seek to balance agility with security, the adoption of these best practices can accelerate the industry wide shift toward proactive security integration, fostering a more resilient cybersecurity ecosystem.

Ultimately, this research contributes to both academia and industry by bridging the gap between DevSecOps theory and practical implementation. By providing SMEs with tailored strategies, the study empowers businesses to scale their security initiatives effectively, positioning security as a catalyst for innovation and growth rather than a development bottleneck.

\section*{Acknowledgment}

Grammarly and ChatGPT assisted with editing, spell checking, and grammar refinement. All content, analysis, and ideas presented in this paper are solely the authors' original work. 

\printbibliography

@INPROCEEDINGS{b1,
  title={Incorporating of Security Methods into the Software Development Lifecycle Process (SDLC)},
  author={Aishwarya, Vakkalagadda and Pediredla, SaiVishal and Radhika, Batchu and Vasanthi, Bhumula and Velmurugan, AK and others},
  booktitle={2023 International Conference on Computer Communication and Informatics (ICCCI)},
  pages={1--4},
  year={2023},
  organization={IEEE}
}

@INPROCEEDINGS{b2,
  author={Alawneh, Muntaha and Abbadi, Imad M.}, 
  booktitle={Ninth Int. Conf. Softw. Defin. Syst. (SDS)},
  title={Expanding {DevSecOps} Practices and Clarifying the Concepts within {Kubernetes} Ecosystem}, 
  year={2022},  
  doi={10.1109/SDS57574.2022.10062874}}

@INPROCEEDINGS{b3,
  author={Aljohani, Mohammad A. and Alqahtani, Sultan S.},
  booktitle={Int. Conf. Smart Comput. Appl. (ICSCA)}, 
  title={A Unified Framework for Automating Software Security Analysis in {DevSecOps}}, 
  year={2023},  
  doi={10.1109/ICSCA57840.2023.10087568}}

@INPROCEEDINGS{b4,
  author={Angermeir, Florian and Voggenreiter, Markus and Moyón, Fabiola and Mendez, Daniel},
  booktitle={ICSE-SEIP 2021},
  title={Enterprise-Driven Open Source Software: A Case Study on Security Automation}, 
  year={2021},
  doi={10.1109/ICSE-SEIP52600.2021.00037}}

@INPROCEEDINGS{b5,
  author={Arseneault, Emily and Boudreau, Daniel and Lien, Jarred and Young, Gregory},
  booktitle={2022 IEEE 29th Annual Software Technology Conference (STC)}, 
  title={Experience-Based Guidelines for Effective Planning \& Management of Software Integration \& Test Activities in the Agile/DevSecOps Environment}, 
   year={2022},
  doi={10.1109/STC55697.2022.00028}}

@inproceedings{b6,
author = {Ashenden, Debi and Ollis, Gail},
title = {Putting the Sec in DevSecOps: Using Social Practice Theory to Improve Secure Software Development},
year = {2021},
doi = {10.1145/3442167.3442178},
booktitle = {New Security Paradigms Workshop (NSPW '20)}}

@ARTICLE{b12,
  author={Díaz, Jessica and Pérez, Jorge E. and Lopez-Peña, Miguel A. and Mena, Gabriel A. and Yagüe, Agustín},
  journal={IEEE Access}, 
  title={Self-Service Cybersecurity Monitoring as Enabler for DevSecOps}, 
  year={2019},
  volume={7},
  doi={10.1109/ACCESS.2019.2930000}}

@INPROCEEDINGS{b13,
  author={Díaz, Oswaldo and Muñoz, Mirna},
  booktitle={6th Int. Conf. Softw. Process Improv. (CIMPS)},
  title={Reinforcing {DevOps} approach with security and risk management: An experience of implementing it in a data center of a {Mexican} organization}, 
  year={2017},
  doi={10.1109/CIMPS.2017.8169957}}

@INPROCEEDINGS{b17,
  author={Göttel, Christian and Kabir-Querrec, Maëlle and Kozhaya, David and Sivanthi, Thanikesavan and Vuković, Ognjen},
  booktitle={IEEE 28th Int. Conf. Emerg. Technol. Factory Autom. (ETFA)}, 
  title={Qualitative Analysis for Validating {IEC} 62443-4-2 Requirements in DevSecOps}, 
  year={2023},
  %pages={1-8},
  doi={10.1109/ETFA54631.2023.10275637}}

@INPROCEEDINGS{b21,
  author={Lee, Jong Seok},
  booktitle={ISSREW 2018}, 
  title={The DevSecOps and Agency Theory}, 
  year={2018},
  doi={10.1109/ISSREW.2018.00013}}

@INPROCEEDINGS{b22,
  author={Mahboob, Jamal and Coffman, Joel},
  booktitle={IEEE 11th Annu. Comput. Commun. Workshop Conf. (CCWC)}, 
  title={A {Kubernetes} {CI/CD} Pipeline with {Asylo} as a Trusted Execution Environment Abstraction Framework}, 
  year={2021},
  doi={10.1109/CCWC51732.2021.9376148}}

@INPROCEEDINGS{b23,
  author={Mao, Runfeng and Zhang, He and Dai, Qiming and Huang, Huang and Rong, Guoping and Shen, Haifeng and Chen, Lianping and Lu, Kaixiang},
  booktitle={IEEE 20th Int. Conf. Softw. Qual. Reliab. Secur. (QRS)}, 
  title={Preliminary Findings about DevSecOps from Grey Literature}, 
  year={2020},
  doi={10.1109/QRS51102.2020.00064}}

@inproceedings{b27,
author = {Morales, Jose Andre and Scanlon, Thomas P. and Volkmann, Aaron and Yankel, Joseph and Yasar, Hasan},
title = {Security impacts of sub-optimal DevSecOps implementations in a highly regulated environment},
year = {2020},
publisher={ACM},
doi = {10.1145/3407023.3409186},
booktitle = {Int. Conf. Availab. Reliab. Secur. (ARES '20)}
}

@inproceedings{b28,
author = {Nadgowda, Shripad and Luan, Laura},
title = {{t}apiserí: Blueprint to modernize DevSecOps for real world},
year = {2021},
doi = {10.1145/3493649.3493655},

booktitle = {7th Int. Workshop Contain. Technol. Contain. Clouds (WoC '21)}
}

@book{creswell2017,
  author    = {John W. Creswell and J. David Creswell},
  title     = {Research Design: Qualitative, Quantitative, \& Mixed Methods Approaches},
  year      = {2017} ,
  publisher = {SAGE Publications}
}

@article{davis1989perceived,
  author={Davis, Fred D.},
  title={Perceived usefulness, perceived ease of use, and user acceptance of information technology},
  journal={MIS Quarterly},
  year={1989},
  volume={13},
  number={3},
  doi={10.2307/249008}
}

@book{rogers2003diffusion,
  author={Rogers, Everett M.},
  title={Diffusion of Innovations},
  year={2003},
  publisher={Free Press},
  pages={319-340}
}

@misc{fowler2006,
author={Fowler, Martin},
title={Continuous Integration},
year={2006},
urldate={2024-03-17},
url={https://martinfowler.com/articles/continuousIntegration.html}}

@misc{centiment,
author={Centiment},
year={2025},
urldate={2025-03-17},
url={https://www.centiment.co/}}

\end{document}